%% file: lqcd-solver.tex
\documentclass[conference]{IEEEtran}
\IEEEoverridecommandlockouts
\usepackage{cite}
\usepackage{amsmath,amssymb,amsfonts}
\usepackage{graphicx}
\usepackage{textcomp}
\usepackage{xcolor}
\usepackage[caption=false,font=footnotesize]{subfig}
\usepackage{fixltx2e}
\usepackage{booktabs}
\usepackage{url}
\usepackage{algorithm}
\usepackage{algpseudocode}
\usepackage{siunitx}
\usepackage{todonotes}
\usepackage{multirow}

 \algnewcommand\And{\textbf{and}}

\usepackage[normalem]{ulem}

\def\BibTeX{{\rm B\kern-.05em{\sc i\kern-.025em b}\kern-.08em
    T\kern-.1667em\lower.7ex\hbox{E}\kern-.125emX}}

\def\Ifp{I_\mathrm{fp}}
\def\Imem{I_\mathrm{mem}}
\def\AI{\mathrm{AI}}
\begin{document}

\title{Performance-Portable Optimization and Analysis of Multiple Right-Hand Sides in a Lattice QCD Solver}

\author{\IEEEauthorblockN{ 
Shiting Long\textsuperscript{1},
Gustavo Ramirez-Hidalgo\textsuperscript{2}, 
Stepan Nassyr\textsuperscript{2}, \\
Jose Jimenez-Merchan\textsuperscript{3}, 
Andreas Frommer\textsuperscript{3},
Dirk Pleiter\textsuperscript{1,4}}
\IEEEauthorblockA{\textit{\textsuperscript{1} KTH Royal Institute of Technology,  Stockholm, Sweden} }
\IEEEauthorblockA{\textit{\textsuperscript{2} Forschungszentrum J\"ulich GmbH, J\"ulich, Germany }}
\IEEEauthorblockA{\textit{\textsuperscript{3} University of Wuppertal,  Wuppertal, Germany}}
\IEEEauthorblockA{\textit{\textsuperscript{4} University of Groningen,  Groningen, Netherlands}}
\IEEEauthorblockA{Email: \textsuperscript{1}{shitingl@kth.se} \textsuperscript{2}{g.ramirez.hidalgo@fz-juelich.de} \textsuperscript{2}{s.nassyr@fz-juelich.de} \\ \textsuperscript{3}{jimenez@math.uni-wuppertal.de}, \textsuperscript{3}{frommer@uni-wuppertal.de} \textsuperscript{1}{pleiter@kth.se}}
}

\maketitle

\begin{abstract}
      Managing the high computational cost of iterative solvers for sparse linear systems is a known challenge in scientific computing. Moreover, scientific applications often face memory bandwidth constraints, making it critical to optimize data locality and enhance the efficiency of data transport. We extend the lattice QCD solver DD-$\alpha$AMG to incorporate multiple right-hand sides (rhs) for both the Wilson-Dirac operator evaluation and the GMRES solver, with and without odd-even preconditioning.  To optimize auto-vectorization, we introduce a flexible interface that supports various data layouts and implement a new data layout for better SIMD utilization. We evaluate our optimizations on both x86 and Arm clusters, demonstrating performance portability with similar speedups. A key contribution of this work is the performance analysis of our optimizations, which reveals the complexity introduced by architectural constraints and compiler behavior. Additionally, we explore different implementations leveraging a new matrix instruction set for Arm called SME and provide an early assessment of its potential benefits.
\end{abstract}


\input{01_introduction}
\input{02_related}
\input{03_methodology}

\input{04_specs}

\input{05_results}
\input{06_SME}
\input{07_conclusions}

\section*{Acknowledgment}
The authors would like to thank the Stony Brook Research Computing and Cyberinfrastructure, and the Institute for Advanced Computational Science at Stony Brook University for access to the innovative high-performance Ookami computing system, which was made possible by a \$5M National Science Foundation grant (\#1927880).
The authors gratefully acknowledge the Gauss Centre for Supercomputing e. V. for funding this project by providing computing time through the John von Neumann Institute for Computing (NIC) on the GCS Supercomputer JUWELS at J\"{u}lich Supercomputing Centre (JSC), under the project with id MUL-TRA.
Furthermore, we want to thank the Open Edge and HPC Initiative for access to an Arm-based development environment through the HAICGU cluster at the Goethe University of Frankfurt. 

Funding for parts of this work has been received from the European Union’s HORIZON MSCA Doctoral Networks programme AQTIVATE, under grant agreement No. 101072344. 
G.R-H. acknowledges financial support from the EoCoE-III project, which has received funding from the European High Performance Computing Joint Undertaking under grant agreement No. 101144014. 

\bibliographystyle{IEEEtran}
\bibliography{lqcd-solver}

\end{document}

%% file: 01_introduction.tex
\section{Introduction}
\label{section:introduction}

Although the computational power of HPC systems has grown rapidly over the years, scientific computations often remain memory-bound due to large datasets and complex memory access patterns. This is because memory technologies struggle to scale at the same rate as advances in computational capabilities, creating a bottleneck in HPC architectures. 

Lattice quantum chromodynamics (LQCD) is a non-perturbative method that employs a discretized 4-dimensional space-time lattice to investigate the strong interaction, a fundamental force governing the interactions between quarks and gluons. LQCD applications are memory-bound, as the computations are relatively lightweight compared to the volume of data that must be moved between memory and processors, making memory bandwidth and latency the limiting factors.
Depending on the LQCD application and the specific workload, a very significant part (often $>90\%$) of the execution time is spent in iterative solvers.

In recent years, there has been a trend of introducing multiple right-hand sides (rhs) in LQCD solvers for performance improvement because 1) matrix-vector operations for a single rhs are replaced by batched operations, increasing the arithmetic intensity, and 2) block Krylov subspace solvers can reduce iterations to convergence by enabling information sharing among multiple rhs. However, the exploration of this rhs blocking lacks a detailed and systematic assessment of its performance. This is especially relevant when such blocking techniques are integrated with algorithmic advances, as the individual impact of each on performance remains unclear. 

As HPC systems evolve, traditional LQCD codes often struggle to adapt to modern hardware capabilities. With the growing diversity of modern CPU architectures, ensuring application portability is essential to accommodate foreseeable hardware changes and maintain performance efficiency. 

In this paper, we focus on optimizing the C++ version~\cite{artur2022ddalphaamg} of the open-source LQCD solver DD-$\alpha$AMG~\cite{frommer2014adaptive}, but we emphasize that it serves as an example with insights that can be broadly applied to other scientific computing applications. 

This work makes the following contributions:
Firstly, we implement rhs blocking in the Wilson-Dirac operator evaluation kernel and integrate it into a batched GMRES solver within DD-$\alpha$AMG. The previous version of the code is constrained by its reliance on 128-bit SSE intrinsics for efficient execution, which not only limits its ability to utilize the meanwhile available 512-bit SIMD instructions on HPC systems but also restricts its portability to Arm-based architectures. We, therefore, perform a code refactorization that takes auto-vectorization into account and introduce a new data layout tailored for loop vectorization, while preserving a compatible data layout for seamless integration with the legacy code. 

Secondly, we perform a comprehensive performance and architectural analysis of the improved portable solver on three very different HPC architectures: JUWELS (x86), Ookami (Arm), and HAICGU (Arm) clusters. Our results show that rhs blocking in kernels achieves a speedup of 10\% to 24\% compared to the original unblocked versions. However, its performance is limited by the effective memory bandwidth, which is sensitive to the choice of blocking size. We also find that although auto-vectorization with optimized data layouts improves SIMD utilization, performance gains are often limited by architectural constraints. Moreover, we observe that a compiler with auto-vectorization may produce markedly different code for varying blocking sizes, complicating performance prediction. Solver-level algorithmic changes may further reduce kernel-level speedups. These findings indicate that rhs blocking is beneficial but constrained, warranting careful application.

Finally, to address the new hardware capabilities of modern CPU architectures, we present a study of Arm's Scalable Matrix Extension (SME) applied to LQCD. We implement an LQCD routine in the Wilson-Dirac operator evaluation with SME and analyze its potential performance using an emulator.

%
    

The paper is structured as follows: Section~\ref{section:related} reviews previous work, Section~\ref{section:methodology} outlines the methodology and implementations, Section~\ref{section:spec} defines the evaluation configurations, Sections~\ref{section:results} and~\ref{section:smeanalysis} present and discuss the results, and Section~\ref{section:conclusions} provides a summary and outlook.

%% file: 02_related.tex
\section{Related Work}
\label{section:related}

The solution of linear systems \( A x_i = y_i \) for \( i = 1, \ldots, b \) rhs can be achieved by expanding a common Krylov subspace generated collectively by all rhs vectors \cite{gutknecht2006-brkylov}. At iteration \( k \), the method has access to the block Krylov subspace
\[
K_k(A, Y) = \mathrm{span} \{ Y, A Y, A^2 Y, \ldots, A^{k-1} Y \},
\]
where \( Y = [ y_1 \mid \ldots \mid y_b ] \) is the matrix of all rhs. This combined subspace contains as subsets the individual Krylov subspaces corresponding to each \( y_i \), enabling simultaneous residual reduction across all systems. 

Early parallel implementations of the methods concerning multiple rhs include block versions of GMRES \cite{Vital1990,doi:10.1137/120883037}, and a block version of BiCGSTAB \cite{block-BiCGSTAB}. To address the stability issues of the latter, Block BiCGR was introduced in \cite{block-BiCGGR}. 
As an alternative to these algorithms, the EigCG method was developed in \cite{eigcg} as a modified Conjugate Gradient (CG) approach that builds an enlarged search space by accumulating a window of past residuals during the CG iterations.

The block solvers have been adopted in physics applications to address specific computational challenges. Forcrand et al.~\cite{forcrand2018RHMC} utilized these methods to reduce variance in force term calculations, while Boyle et al. \cite{boyle-mg-block-preconditioed} applied them to domain wall fermion systems. They were also successfully implemented in the context of LQCD for the Wilson-Dirac operator in~\cite{block-krylov-lqcd}. 

Recent work of block solvers has focused on enhancing their efficiency for modern HPC systems. An efficient implementation of block Krylov methods on GPUs was presented in \cite{block-gpu}. 
Birk et al. \cite{mult-shifts-mult-rhs} incorporate simultaneous inversion of multiple rhs and shifted systems by employing block Krylov solvers in DD-$\alpha$AMG. 
%

LQCD has always been an area of research that was strongly dependent on the efficient use of HPC resources.
This resulted in various performance analysis and modeling efforts, Bilardi et al.~\cite{doi:10.1007/11602569_41} and Bauer et al.~\cite{doi:10.1109/CCGrid.2012.123} being early examples.
The former is of particular interest for this work as it introduces a method for systematic analysis of the data transfers that are typically limiting the performance.
The need for performance has also been a driver for exploring new processor architectures.
The A64FX processor, which is also considered in this work, has been of immediate great interest for LQCD applications due to its high memory bandwidth~\cite{doi:10.1109/CLUSTER.2018.00079}.
Alappat et al.~\cite{doi:10.1109/PMBS51919.2020.00006} present a performance model for an application on A64FX that has some similarities with the application considered in this work.

Arm's recent instruction set architecture (ISA) extension SME has been a topic of only a few studies.
Remke and Beuer~\cite{doi:10.1109/SCW63240.2024.00185} explored relatively simple kernels on a recent Apple processor.
Hübner et al.~\cite{doi:10.48550/arXiv.2502.05317} extended this to more complex HPC applications.
Zhang et al.~\cite{doi:10.1109/IPDPS57955.2024.00088} introduced an optimized implementation of FFT using SME instructions.
Other vendors also introduced instructions supporting matrix or tensor operations.
One example is Intel's Advanced Matrix Extensions (AMX) (see, e.g., \cite{doi:10.1145/3633332} for an evaluation).
Since it does not support double-precision floating-point operations, it is not considered here.

%% file: 03_methodology.tex
\section{Optimizations for DD-$\alpha$AMG}
\label{section:methodology}

This section describes our implementations for optimizing the DD-$\alpha$AMG code. We start with introducing rhs blocking in Section~\ref{subsec:alg}, followed by a redesign of the data layout for efficient SIMD utilization in Section~\ref{subsec:datalayout}. Lastly, we detail SME implementations for matrix operations in Section~\ref{subsec:smekernel}.

\subsection{Multiple RHS in DD-$\alpha$AMG}
\label{subsec:alg}


DD-$\alpha$AMG is an aggregation-based adaptive algebraic multigrid method using domain decomposition to solve LQCD systems with the Wilson-Dirac operator. All but the coarsest levels use flexible GMRES with multigrid preconditioning, while the last applies odd-even GMRES. Since the finest level dominates computational cost, this paper focuses on optimizing the GMRES solver on that level and omits multigrid preconditioning to avoid added complexity.

\subsubsection{Wilson-Dirac operator evaluation}

A significant portion of the computational time in LQCD solvers is spent on the evaluation of the Wilson-Dirac operator, making the part critical for optimization. We give the formulation of the discretized Wilson-Dirac operator as follows:
\[
\begin{split}
    D_W \psi(x) = \frac{4+m_0}{a} \psi(x) - \frac{1}{a} \sum_{\mu=0}^{3} \left( \pi^-_\mu \otimes U_\mu(x) \psi(x + \hat{\mu})\right.\\
\left. + \pi^+_\mu \otimes U_\mu^\dagger(x - \hat{\mu}) \psi(x - \hat{\mu}) \right),
\end{split}
\]
where $m_0$ denotes the mass parameter, $\psi$ denotes the input vector, and $x$ denotes a lattice site. $\mu=0,\ldots,3$ is the dimension index, $\hat{\mu}$ is a shift vector. The lattice spacing $a$ is 1 in this work. $\pi^\pm_\mu$ are constant projectors, and $U$ represents the gauge configuration. The formulation has two components: the self-coupling term (the first term), which accounts for site-local interactions, and the hopping term (the second term), which describes interactions between neighboring sites. To reduce the discretization error, a clover term $C(x)$ is introduced, and the improved Wilson-Dirac operator is written as 
\[D\psi(x) = D_W\psi(x) - C(x)\psi(x).\]
We present the algorithm for the Wilson-Dirac operator evaluation \cite{rottmann2016adaptive} in Algorithm~\ref{alg:wilson-dirac}.

\begin{algorithm}[btp]
 \caption{Wilson-Dirac Operator Evaluation}\label{alg:wilson-dirac}
\begin{algorithmic}[1]
 \Require $\psi$
 \Ensure  $\eta = D \psi$
   \State $\forall x$, Evaluate $\eta(x) \leftarrow \left(4 + m_0\right)\psi(x)-C(x)\psi(x)$
   \State $\forall x, \forall \mu$, Evaluate $\lambda_\mu(x) \leftarrow \pi^{-}_{\mu}\otimes I_3\psi(x)$\;
   \State $\forall \mu$, $\forall -\hat{\mu}$ halo site $x'$, Send $\lambda_\mu(x')$ to $-\hat{\mu}$ neighbor of $x'$\;
  \State  $\forall x, \forall \mu$, Evaluate $\chi_\mu(x+\hat{\mu}) \leftarrow \pi^+_\mu \otimes U_\mu^\dagger(x) \psi(x)$\;
  \State  $\forall \mu$, $\forall +\hat{\mu}$ halo site $x''$, Send $\chi_\mu(x''+\hat{\mu})$ to $+\hat{\mu}$ neighbor of $x''$\;
   \State $\forall \mu$, $\forall +\hat{\mu}$ halo site $x''$, Receive $\lambda_\mu(x''+\hat{\mu})$ from $+\hat{\mu}$ neighbor\;
  \State  $\forall x, \forall \mu$, Evaluate $\eta(x) \leftarrow \eta(x) - [I_4\otimes U_\mu(x)]\lambda_\mu(x+\hat{\mu})$\;
   \State $\forall \mu$, $\forall -\hat{\mu}$ halo site $x'$, Receive $\chi_\mu(x')$ from $-\hat{\mu}$ neighbor\;
   \State $\forall x, \forall \mu$, Evaluate $\eta(x) \leftarrow \eta(x) - \chi_\mu(x)$\;
 \end{algorithmic} 
 \end{algorithm}

In the DD-$\alpha$AMG code, each lattice site is identified by four coordinates corresponding to the four dimensions of the lattice. To assign a unique site number, these coordinates are mapped into a single integer using a positional numbering system where the stride for each dimension is defined by its size. Suppose the lattice has size $N_0 \times N_1 \times N_2 \times N_3$, a site coordinate $(x_0,x_1,x_2,x_3)$ can be mapped to a site number by calculating 
\[
x_0\times N_1 N_2  N_3 + x_1\times N_1 N_2 + x_2\times N_1 + x_3. 
\]

The array $\psi$ is one-dimensional and iterates through the lattice sites in ascending order according to their unique site numbers. We write $\psi = [\psi(1) \;\psi(2)\; \ldots \;\psi(N_L)]^T$, where $N_L$ denotes the total number of lattice sites. For any site $x$, $\psi(x)\in\mathbb{C}^{12}$, indicating that
\[
\begin{split}
    \psi = [&\psi_1(1)\; \psi_2(1) \;\ldots \;\psi_{12}(1) \;\psi_1(2)\; \ldots\\
    &\ldots \;\psi_{12}(2) \;\ldots \;\psi_1(N_L)\; \ldots\; \psi_{12}(N_L)]^T.
\end{split}
\]

The gauge configuration $U$ is stored in a two-dimensional array with lattice sites and the lattice spatial dimensions as indices. Note that for any array dimension representing lattice sites, the storage order follows the same site numbering convention as discussed above. Thus, we write
\[
U_\mu=[U_\mu(1) \; U_\mu(2) \; \ldots \; U_\mu(N_L)]^T, \mu=0\ldots3. 
\]
Each $U_\mu(x)$ is a $3\times 3$ matrix, stored in row-major order, i.e., $U_\mu(x)\in\mathbb{C}^{3\times3}$. Its applications in lines 4 and 7 of Algorithm~\ref{alg:wilson-dirac} can be interpreted as applying $U_\mu(x)$ to vectors of length 3 simultaneously. 

The clover term is block-diagonal, with each block being Hermitian, thus, $C(x)\in\mathbb{C}^{42}$. The projectors $\pi^\pm_\mu$ are block off-diagonal and anti-Hermitian matrices, enabling the generation of $\lambda_\mu$ and $\chi_\mu$ such that only 6 elements are needed per lattice site, i.e., $\lambda_\mu(x), \chi_\mu(x)\in\mathbb{C}^6$. In practice, this method is applied in lines 2 and 4 of Algorithm~\ref{alg:wilson-dirac}. When the data are written to $\eta(x)$ (lines 7 and 9), additional linear combinations are performed to restore the data to length 12.

Introducing multiple rhs to the Wilson-Dirac operator evaluation, i.e., $\psi = [\psi^{(1)} | \psi^{(2)} | \cdots | \psi^{(b)}]$, where the superscript denotes rhs numbering, we consider a block of vectors in form $v = [v(1) \;v(2) \;\ldots\; v(N_L)]^T$ as a block-vector (matrix):
\begin{equation}
\label{eq:block}
    V = 
\begin{bmatrix}
    v^{(1)}(1)& v^{(2)}(1) &\cdots  &v^{(b)}(1)\\
    v^{(1)}(2)& v^{(2)}(2) &\cdots &v^{(b)}(2)\\
    \vdots &  \vdots &  & \vdots\\
    v^{(1)}(N_L)& v^{(2)}(N_L) &\cdots  &v^{(b)}(N_L)\\
\end{bmatrix}.
\end{equation}
To solve the system with $\psi$ stored in a column-major order, one can apply Algorithm~\ref{alg:wilson-dirac} iteratively with each column as an rhs. However, this method does not improve data locality. A more efficient approach is to use a row-major storage, which then integrates the loop over rhs into a loop over lattice site $x$, e.g., line 4 in the algorithm can be rewritten as
\[
    \forall x, \forall \mu, \forall i=1\ldots b, \text{Evaluate } \chi^{(i)}_\mu(x+\hat{\mu}) \leftarrow \pi^+_\mu \otimes U_\mu^\dagger(x) \psi^{(i)}(x).\;
\]
Note that $\lambda_\mu$, $\chi_\mu$ and $\eta$ now also have an additional dimension of size $b$. This approach increases the arithmetic intensity of Algorithm~\ref{alg:wilson-dirac} by reducing total memory access, as neither the clover term nor the gauge configuration is revisited. Disregarding caching effects, the arithmetic intensity of Algorithm~\ref{alg:wilson-dirac} is
\begin{equation}
\label{eq:arithmetic}
  \AI(b) =  \frac{\Ifp(b)}{\Imem(b)} = \frac{2574\,b\,\mathrm{Flop}}{(168\,b + 114)\,16\,\mathrm{Byte}},
\end{equation}
where $b$ denotes the blocking size, $16$ the size of double-precision complex numbers. $\Ifp$ and $\Imem$ capture the number of floating-point operations and the amount of data exchanged with the memory, respectively.

\subsubsection{Batched GMRES Solver}

So far, we have introduced rhs blocking in an LQCD computational kernel, i.e., the Wilson-Dirac operator evaluation. Next, we detail an iterative solver incorporating the kernel to study rhs blocking within a full LQCD solution process.


To support the blocked evaluation of the Wilson-Dirac operator, we implement a batched GMRES solver (Algorithm~\ref{alg:bgmres}), which processes multiple rhs in each iteration. The blocked evaluation of the Wilson-Dirac operator is called in line 2 of the algorithm, where $D\psi^{(i)}$ is computed for each $i=1\ldots b$. In a complete solve, the overhead of rearranging data layout for better locality is negligible over hundreds of iterations. Even with column-major storage, converting to row-major before solving and reverting afterward remains practical.

\begin{algorithm}[btp]
\caption{Batched GMRES}\label{alg:bgmres}
\begin{algorithmic}[1]
\Require {$D, \eta$, tolerance $\epsilon$}
\Ensure {$\forall i=1\ldots b, D\psi^{(i)}=\eta^{(i)}$}
\For{$rn= 0; rn<restartNum\;\&\&\;!finish; rn++$}

   \State $\forall i$, Compute $r^{(i)}=\eta^{(i)}-D\psi^{(i)}$, 
   
   $norm^{(i)} = \|r^{(i)}\|_2, V[0]^{(i)} = r^{(i)}/norm^{(i)}$\;
   \For{$rl=0; rl < restartLen \;\&\& \;!finish; rl++$}
        \State $\forall i$, Do one step of Arnoldi with $V[j]^{(i)}$\;
        \State $\forall i$, Update $H^{(i)}, \gamma^{(i)}$, 
        
        \;\;\;\;\;Compute $relnorm^{(i)}=\gamma[j+1]^{(i)}/\|\eta^{(i)}\|_2$\;
        \If{$max(relnorm)< \epsilon$}
            \State $finished = 1$\;
        \EndIf
   \EndFor
   
\State $\forall i$, Compute $\psi^{(i)} = \psi^{(i)} + V^{(i)}*y^{(i)}$,
   
where $\| \gamma^{(i)}-H^{(i)}y^{(i)}\|_2 \to \min$
\EndFor
\end{algorithmic}
\end{algorithm}

Instead of using a block Krylov subspace solver, we use a batched version, where each rhs is solved independently. This approach benefits only from machine-level performance gains due to a more favorable memory access pattern. In contrast, block Krylov subspace solvers provide an additional algorithmic performance advantage, making it difficult to isolate and assess the true impact of each factor.


Odd-even preconditioning is a common technique used in LQCD, which typically reduces the number of iterations by a factor of 2 to 3 \cite{osborn2010multigrid}. It splits the lattice sites into \textit{odd} and \textit{even} sites based on their positions, transforming the linear system with $D$ into a system with Schur complement half its size. We incorporate odd-even preconditioning as an optional feature for batched GMRES, extending its application to the fine-grid level within DD-$\alpha$AMG. 



\subsection{Data Layouts for Multiple RHS}
\label{subsec:datalayout}

The data layout rearrangement for blocking in DD-$\alpha$AMG is more complicated than what has been discussed above. A row from the matrix in Equation~\ref{eq:block} can be expanded as
\begin{equation}
\label{eq:rowv}
\text{Row}_x(V) = 
    \begin{bmatrix}
    v^{(1)}_1(x) & v^{(2)}_1(x) & \cdots & v^{(b)}_1(x)\\
    v^{(1)}_2(x) & v^{(2)}_2(x) & \cdots & v^{(b)}_2(x)\\
    \vdots &  \vdots &  & \vdots\\
    v^{(1)}_{s}(x) & v^{(2)}_{s}(x) & \cdots & v^{(b)}_{s}(x)\\
\end{bmatrix},
\end{equation}
where $s$ denotes the number of elements associated with $v(x)$. This matrix can be stored either in column-major (Layout 1) or row-major (Layout 2) order, both are supported and switchable via an interface in our implementation. 

To compare the two layouts, we use a matrix-vector multiplication function as an example. We take the operation in line 7 of Algorithm~\ref{alg:wilson-dirac} for demonstration, in which a $3\times3$ matrix ($U_\mu(x)$) is applied to vectors of length 3 (taken from $ \lambda_\mu(x+\hat{\mu})$). For simplicity, we denote the $3\times3$ matrix as $A$ and let $M$ represent the matrix in Equation~\ref{eq:rowv}, with $s=3$.

\begin{figure}[btp]%
    \centering
    \subfloat[\centering Layout 1]{{\includegraphics[width=0.4\textwidth]{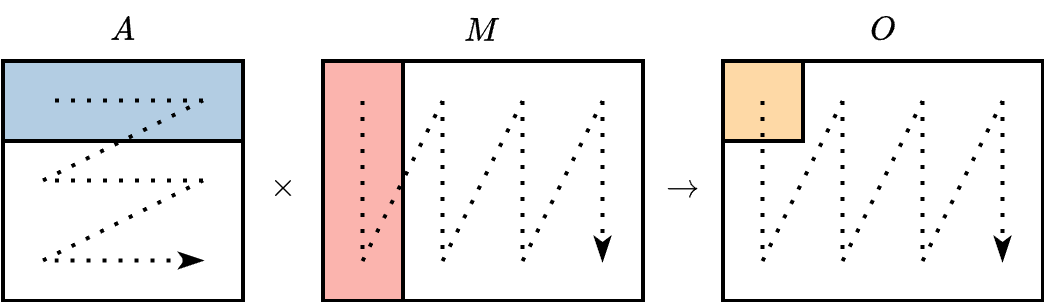} }}%
    \qquad
    \subfloat[\centering Layout 2]{{\includegraphics[width=0.4\textwidth]{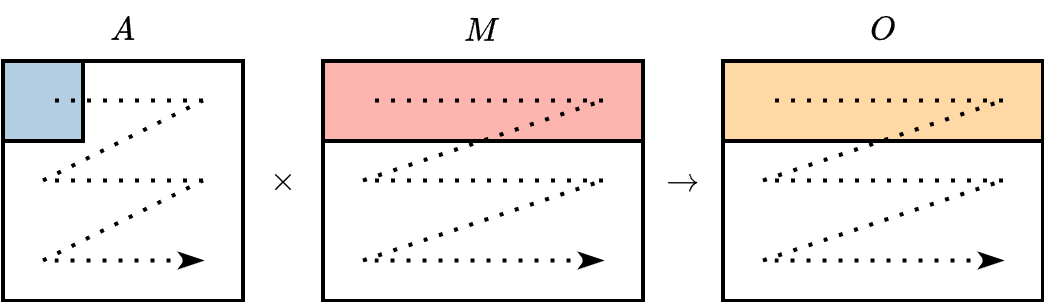} }}%
    \caption{Example of matrix operation using two data layouts with $b=4$, where dotted lines and arrows represent the storage order in memory.}%
    \label{fig:datalayout}%
\end{figure}

For Layout 1, $M$ is stored in a column-major order. This layout requires minimal modification of the original code, as the function for applying a matrix to a vector can be reused. We can perform $A\,m^{(1)}$ until $A\,m^{(b)}$ in the loop over the blocking size $b$. This layout, however, makes it difficult to fully utilize SIMD instructions. As depicted in Fig.~\ref{fig:datalayout}(a), a sequential load of $A$ results in row-wise data, and the load of $M$ results in column-wise data; a multiplication of the two results in a scalar data. For effective SIMD use, one can perform a scatter load of either $A$ or $M$, thereby introducing overhead.

To overcome this problem, we can store $M$ in a row-major order as in Layout 2. As illustrated in Fig.~\ref{fig:datalayout}(b), one can use a masked load of the scalar in $A$ and then broadcast it to all lanes in a SIMD register. A sequential load of $M$ gives row-wise data, and the multiplication also results in row-wise data for output. This makes Layout 2 more efficient for SIMD processing compared to Layout 1.

In the past, in case of complex arithmetic on processor architectures supporting SIMD instructions, it has often been beneficial to store real and imaginary values in different arrays to avoid the need for SIMD-lane permutations.
Modern SIMD instruction set architectures like AVX and SVE come with good support for complex arithmetics (examples are provided in Section~\ref{subsec:smekernel}), such that it is more beneficial to work with arrays of complex numbers as it improves data locality. Thus, we maintain the natural complex number storage format rather than separating components.

\subsection{SME Implementations}
\label{subsec:smekernel}
While LQCD applications involve large-scale linear systems, the resulting matrices are predominantly sparse. Therefore, employing SME for the entire linear system in DD-$\alpha$AMG is inadvisable as it is computationally inefficient. We decide to implement a matrix-matrix multiplication function realizing $AM = O$ as discussed in Section~\ref{subsec:datalayout} with SME. Although lightweight, this function accounts for significant computational time during the Wilson-Dirac operator evaluation, making it non-negligible to the overall LQCD solve.

The ZA tile is a fundamental architectural component in SME, which is a large two-dimensional array register with dimensions based on the scalable vector length (SVL). Since $A$ has a size of $3\times3$ and $M$ a size of $3\times b$, the code would underutilize a ZA tile. For example, considering a 256-bit or 512-bit SVL, the size of 3 complex doubles ($3\times128$ bits) would leave lanes of size 128-bit in the ZA tile unused.

The key instruction in SME is \verb|fmopa|, which calculates the outer product of two SVL vectors and accumulates. Since $O$ can be calculated by repeated addition of $A_{i^{th}col} \otimes M_{i^{th}row}$ for $i=1\ldots3$, the operation $AM=O$ can be directly implemented using the \verb|fmopa| instructions. 

The practical computation is, however, more complicated than a simple application of \verb|fmopa| due to complex number arithmetic. A multiplication of two complex numbers requires four real number multiplications. These multiplications require a special vector loading mechanism in SVE/SME as the real and imaginary values are stored in an interleaved format. The SVE/SME ISAs provide the deinterleaving structured load instruction \verb|ld2d|, which can place real and imaginary values into two vector registers. An interleaving store \verb|st2d| is also available for storing data from two separate vector registers to memory. For sufficiently large vectors or matrices, the two instructions enable an efficient workflow: 1) load operands with \verb|ld2d|, 2) compute the real and imaginary partial products separately, then 3) store results using \verb|st2d|. 

Given that both $A$ and $M$ contain at least one small dimension, we implement an additional optimization by only deinterleaving one of the operands. This approach allows more active lanes in the vector registers and the ZA tile, making more efficient use of available hardware resources. However, this approach requires storing a copy of the interleaved operand with the real and imaginary values swapped and the imaginary values negated, introducing two additional instructions \verb|revd| and \verb|fneg|.

\begin{figure}[btp]%
    \centering
    \subfloat[\centering Step 1]{{\includegraphics[width=0.46\textwidth]{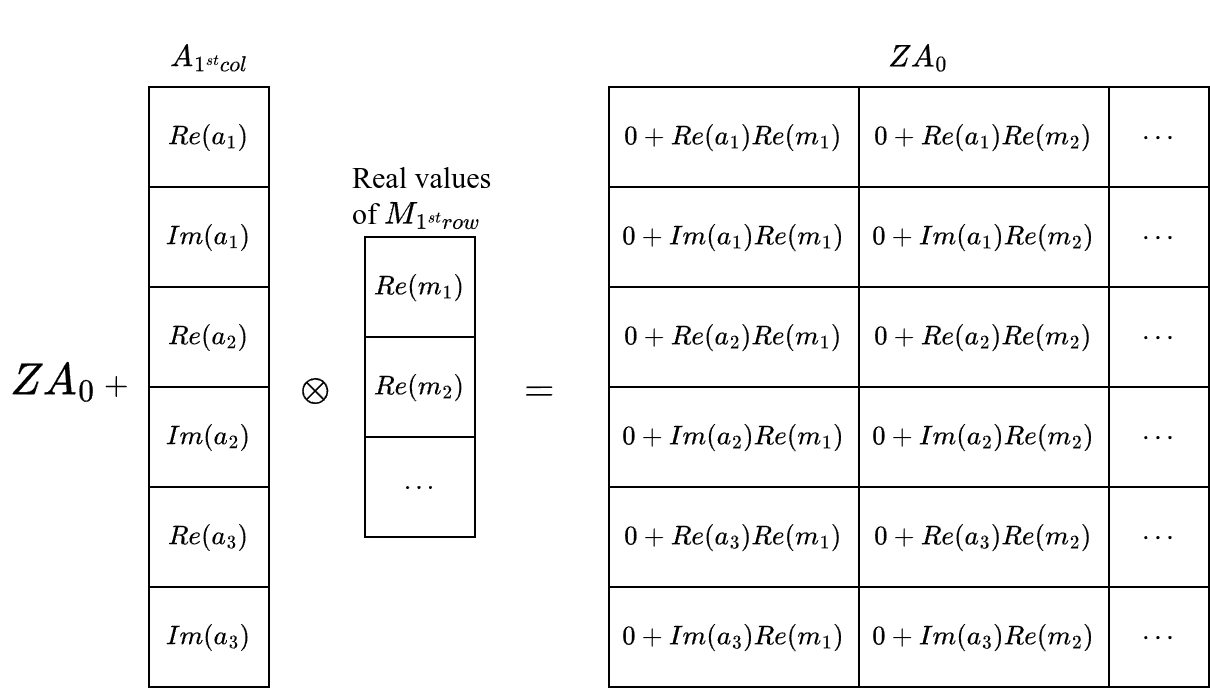} }}%
    \qquad
    \subfloat[\centering Step 2]{{\includegraphics[width=0.46\textwidth]{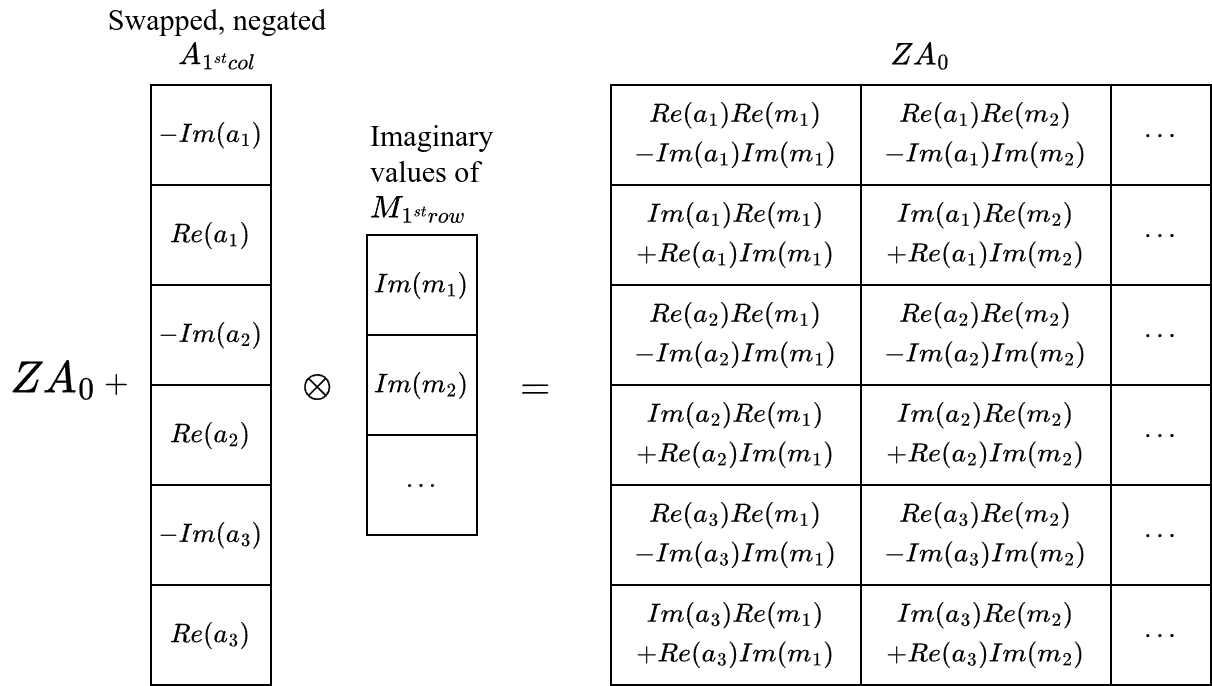} }}%
    \caption{Demonstration of $A_{1^{st}col} \otimes M_{1^{st}row}$ using SME.}%
    \label{fig:sme}%
\end{figure}

While Layout 2 naturally aligns with the memory access pattern of $M$, we further optimize performance by transposing $A$ to guarantee fully sequential access in the multiplication. We therefore adopt a column-major store for $A$ and a row-major store for $M$. We showcase the $A_{1^{st}col} \otimes M_{1^{st}row}$ operation in Fig.~\ref{fig:sme}, which requires two \verb|fmopa| instructions executed in sequential steps. We denote the ZA tile used by $ZA_0$ and initialize it to zero. The unused lanes on the ZA tile are omitted in the figure. In step 1, $A_{1^{st}col}$ can be sequentially loaded, but the real and imaginary values of $M_{1^{st}row}$ must be loaded using the deinterleaving instruction \verb|ld2d| into two separate vector registers. The vector register containing the imaginary values of $M_{1^{st}row}$ is subsequently used for the operation in Step 2. To compute the swapped and negated $A_{1^{st}col}$ for Step 2, we use the \verb|revd| and \verb|fneg| instructions as described above. 

After the two steps, the data stored in $ZA_0$ are
\[
    \begin{bmatrix}
        Re(a_{1}m_{1}) & Re(a_{1}m_{2}) & \ldots\\
        Im(a_{1}m_{1}) & Im(a_{1}m_{2}) & \ldots\\
        Re(a_{2}m_{1}) & Re(a_{2}m_{2}) & \ldots\\
        Im(a_{2}m_{1}) & Im(a_{2}m_{2}) & \ldots\\
        Re(a_{3}m_{1}) & Re(a_{3}m_{2}) & \ldots\\
        Im(a_{3}m_{1}) & Im(a_{3}m_{2}) & \ldots\\
    \end{bmatrix},
\]
which are exactly the complete results of $A_{1^{st}col} \otimes M_{1^{st}row}$ in complex numbers. The $AM=O$ operation requires iterating both steps for all $i= 1\ldots3$.

We implement two SME variants to perform $AM=O$, differing in which one of the input operands is left interleaved and requires the negation and swap operations. The first variant (neg-A) applies the negation and swap operations to elements of $A$, as illustrated in Fig.~\ref{fig:sme}. The second variant (neg-M) instead performs the operations on elements of $M$. In this variant, we sequentially load $M_{i^{th}row}$, but load the real and imaginary values of $A_{i^{th}col}$ separately in the outer product computations. This variant is expected to be suboptimal for a large blocking size $b$, as the required number of negation and swap operations scales with the row dimension of $M$. Meanwhile, columns of $A$ would further underutilize the ZA tile, as only $3\times64$ bits in a dimension are used.

%% file: 04_specs.tex
\section{Test Specifications}
\label{section:spec}

\subsection{Test Platforms}
\label{subsec:platform}
\begin{table*}[btp]
    \centering
        \caption{Node-level hardware and software configurations for benchmarking platforms evaluated for optimized DD-$\alpha$AMG.}
    \begin{tabular}{l @{\hspace{0.6cm}} l @{\hspace{0.6cm}} l @{\hspace{0.6cm}} l}  
\toprule

& JUWELS & Ookami  & HAICGU  \\
\midrule
Architecture & x86 & Arm & Arm\\
Processor  &   2$\times$ Intel Xeon Platinum 8168   & A64FX   &  2$\times$ Kunpeng 920    \\
Number of Cores &  $2\times 24$ & 48 & $2\times 64$\\
CPU Frequency   &   \qty{2.7}{\giga\hertz}   & \qty{1.8}{\giga\hertz}  &  \qty{2.6}{\giga\hertz}  \\
L1 Data Cache & \qty{32}{\kibi\byte} per core  & \qty{64}{\kibi\byte} per core & \qty{64}{\kibi\byte} per core\\
L2 Cache & \qty{1}{\mebi\byte} per core & 4$\times$\qty{8}{\mebi\byte} & \qty{512}{\kibi\byte} per core\\
L3 Cache & \qty{33}{\mebi\byte} per CPU & none & \qty{32}{\mebi\byte} per CPU\\
SIMD Width & \qty{512}{\bit} & \qty{512}{\bit} & \qty{128}{\bit} \\
Memory & \qty{96}{\gibi\byte} & \qty{32}{\gibi\byte} & \qty{128}{\gibi\byte} \\
Theor.~Mem.~Bandwidth   &  \qty{256}{\giga\byte\per\second}   &  \qty{922}{\giga\byte\per\second}   &  \qty{341}{\giga\byte\per\second}   \\
STREAM Triad Bandwidth & \qty{155}{\giga\byte\per\second} (48 cores)   &  \qty{619}{\giga\byte\per\second} (24 cores)  & \qty{218}{\giga\byte\per\second} (128 cores)\\
STREAM Copy Bandwidth & \qty{133}{\giga\byte\per\second} (48 cores) & \qty{540}{\giga\byte\per\second} (24 cores) & \qty{214}{\giga\byte\per\second} (128 cores) \\
Network Bandwidth & \qty{100}{\giga\bit\per\second} & \qty{100}{\giga\bit\per\second} & \qty{100}{\giga\bit\per\second} \\
Compiler Used & GCC v13.3.0 & GCC v13.2.0 & GCC v14.1.0 \\
Compiler Flags for SIMD & \texttt{-mavx512f} & \texttt{-march=armv8.2-a+sve} & \texttt{-march=armv8.2-a+simd}\\
\midrule[\heavyrulewidth]
\bottomrule
\end{tabular}
\label{tab:spec}
\end{table*}

To evaluate our implementations in DD-$\alpha$AMG, we execute the application on three distinct platforms: JUWELS (x86), Ookami~\cite{doi:10.48550/arXiv.2311.04259} (Arm), and HAICGU (Arm). The implementations include rhs blocking with two data layouts in both the Wilson-Dirac operater evaluation (the kernel) and batched GMRES (the solver). 

Table~\ref{tab:spec} lists the hardware and software used for the evaluation in Section~\ref{section:results}. In addition to the theoretical memory bandwidth, we employ the STREAM benchmark \cite{mccalpin1995memory} v5.10 to measure the peak attainable memory bandwidths and the number of cores required to reach them. The choice of using STREAM benchmark is to establish an empirical performance ceiling that is more accurate than the theoretical memory bus limit. On many architectures, STREAM Triad benchmark achieves the highest memory bandwidth. However, the Triad bandwidth may still be optimistic for write-intensive workloads, especially on architectures with asymmetric read and write bandwidths. To account for such cases, we also include the STREAM Copy bandwidths in Table~\ref{tab:spec}.

In the absence of available HPC-grade hardware SME implementations, we use an emulator for SME evaluation. The use of the cycle-level simulator gem5 \cite{binkert2011gem5} was considered, but the SME implementation in gem5 is not mature enough for performance analysis as of this writing. Therefore, we turn to the emulator QEMU \cite{bellard2005qemu} v9.2.50 together with its plugin \textit{insn} for SME evaluation in Section~\ref{section:smeanalysis}. 

\subsection{Test Configurations}

\subsubsection{Empirical performance Evaluation}
We evaluate two lattice configurations with sizes: 1) $128\times64^3$, typical in LQCD simulations~\cite{doi:10.22323/1.430.0203}, and 2) $64\times16^3$, suited for single-node runs. Both lattices are used in the kernel evaluation in Section~\ref{subsec:wdeval}, while only the $128\times64^3$ lattice is used to exemplify the solver evaluation in Section~\ref{subsec:bgmreseval}.

We use GCC auto-vectorization (enabled with \texttt{-O3}) and additional platform-specific compiler flags listed in Table~\ref{tab:spec} for SIMD optimization. Parallelism is achieved using MPI from the existing DD-$\alpha$AMG code. Additionally, OpenMP support is implemented for the Wilson-Dirac operator evaluation. This is necessary to increase memory-level parallelism and leverage the high-bandwidth memory (HBM2) architecture used on Ookami.

Given the higher memory bandwidth of the A64FX nodes and given the dual-socket architecture of the JUWELS nodes, we use twice as many JUWELS nodes (256) as Ookami nodes (128) to ensure a fair performance comparison for multi-node evaluations. Due to hardware limitations (28 nodes), HAICGU cannot accommodate the $128\times 64^3$ lattice configuration. We therefore exclude multi-node HAICGU results in this paper.  

For both multi-node and single-node evaluations, JUWELS and Ookami use the same node-level parallelism settings. For JUWELS, each node is configured with 16 MPI ranks. The decomposition of the lattice requires the number of MPI ranks to be a power of two, so increasing beyond this (e.g., to 32 ranks) would hinder efficient multithreading. While reducing MPI ranks per node allows more threads per rank, we observe suboptimal performance in practice. Thus, we configure 16 MPI ranks per node with 2 threads each, as additional threads yield diminishing returns on JUWELS. Similarly for Ookami, we configure 16 MPI ranks per node, but each with 3 threads to maximize core utilization. The core count of HAICGU, being a power of two, offers architectural advantages. Through empirical testing, we determined the optimal configuration uses 32 MPI ranks with 4 threads per rank on HAICGU. The MPI and OpenMP affinities are carefully configured to ensure an optimal distribution of processes and threads within each node on all platforms.

\subsubsection{Emulated Performance Evaluation of SME}
\label{subsubsec:emu}

We implement a benchmark driver to execute the matrix-matrix multiplication function as discussed in Section~\ref{subsec:smekernel}. The driver supports four variants: a BLIS~\cite{van2016blis} v1.1 implementation of the
ZGEMM routine (function \texttt{cblas\_zgemm}), an auto-vectorized SVE code using GCC v14.2.0 and two of our SME implementations (neg-A and neg-M). To ensure accurate instruction counts, the driver iterates over the region of interest (ROI) multiple times.



We also write a custom performance modeling script that 1) executes the driver via QEMU for two distinct blocking sizes ($b_1, b_2$); 2) computes the instruction-count difference between the two runs; and 3) applies instruction-specific cost weights to derive the total computational cost difference.

This methodology intentionally excludes program overheads, focusing exclusively on isolating and quantifying the computation cost differential attributable to the change in the blocking size interval $|b_1-b_2|$. That is, to evaluate the emulated performance, we run the modeling script to count computation-relevant instructions and apply their associated costs. This method allows us to compare the variants and estimate potential computational speedup under stall-free execution, without taking any memory-related bottlenecks into account. 


%% file: 05_results.tex
\section{Empirical Performance Evaluation}
\label{section:results}





\subsection{Wilson-Dirac Operator Evaluation}
\label{subsec:wdeval}

\subsubsection{Multi-Node Performance}

\begin{figure}[btp]
\vspace{-2.5mm}
\centering
  \includegraphics[width=.46\textwidth]{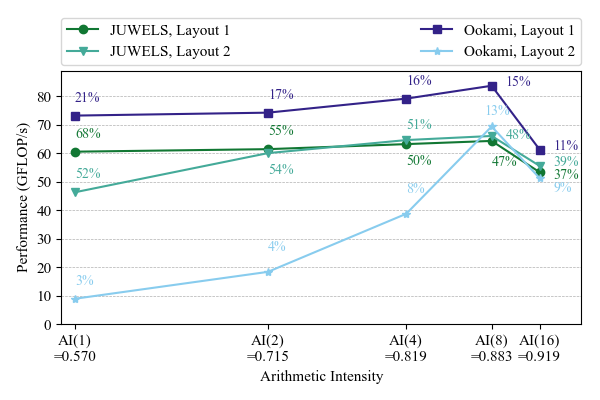}
  \vspace{-3mm}
  \caption{Performance per node scaling with arithmetic intensity in the Wilson-Dirac operator evaluation for the $128\times64^3$ lattice on JUWELS and Ookami, where the percentage labels denote architectural efficiencies.}
  \label{fig:dirac1}
\end{figure}

We use the roofline model to present the multi-node performance in Fig.~\ref{fig:dirac1}, in which $\AI(b)$ denotes the arithmetic intensity for the blocking size $b$; cf.\ Equation~\ref{eq:arithmetic}. Since the application is memory-bound, its performance ceiling depends on the memory bandwidth and the arithmetic intensity. The ceiling is also constrained by factors that affect the theoretical performance calculated based on manufacturer-specified hardware information. The memory bandwidth measured by the STREAM Triad benchmark provides a more realistic upper bound on the attainable memory bandwidth. Thus, we define the ceiling as 
\begin{equation}
    \text{Theor. HW. Perf.}(b) = \text{STREAM Triad Bandwidth} \times \AI(b).
\end{equation}
We further use a metric \textit{architectural efficiency}~\cite{pennycook2019implications} to quantify the measured performance as a percentage of the performance ceiling. The metric is defined as   
\begin{equation}
    \text{Arch. eff.} = \frac{\text{Meas. Perf.}}{\text{Theor. HW. Perf.}}, 
\end{equation}
and the corresponding values are reported for all data points in Fig.~\ref{fig:dirac1}.

Since the original unblocked and not optimized implementation in DD-$\alpha$AMG is equivalent to Layout 1 with blocking size $b=1$, it can be seen from Fig.~\ref{fig:dirac1} that the best-performing rhs blocking variant (Layout 2) achieves up to a 10\% performance gain over the original on JUWELS. The improvement increases to 15\% on Ookami, though Layout 1 is the best-performing rhs blocking variant there.

\subsubsection{Single-node Performance}

\begin{figure}[btp]
  \vspace{-2.5mm}
  \centering
  \includegraphics[width=0.46\textwidth]{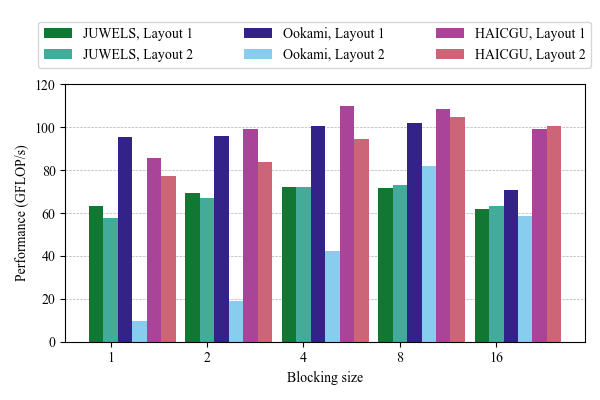}
  \vspace{-3mm}
  \caption{Single-node performance comparison of the Wilson-Dirac operator evaluation for the $64\times 16^3$ lattice across JUWELS, Ookami, and HAICGU.}
  \label{fig:comp}
\end{figure}

\begin{table}[btp]
    \centering
        \caption{Architectural efficiency of the Wilson-Dirac operator evaluation on JUWELS, Ookami, and HAICGU.}

\begin{tabular}{l@{\hspace{0.2cm}} c@{\hspace{0.2cm}} c@{\hspace{0.2cm}} c@{\hspace{0.2cm}} c@{\hspace{0.2cm}}  c@{\hspace{0.2cm}} c@{\hspace{0.2cm}} c@{\hspace{0.2cm}} c@{\hspace{0.2cm}} c@{\hspace{0.2cm}} c@{\hspace{0.2cm}}}
    \toprule
     $b=$& \multicolumn{2}{c}{1} & \multicolumn{2}{c}{2} & \multicolumn{2}{c}{4} 
      & \multicolumn{2}{c}{8} & \multicolumn{2}{c}{16} \\
    \cmidrule(lr){2-3}
    \cmidrule(lr){4-5}
    \cmidrule(lr){6-7}
    \cmidrule(lr){8-9}
    \cmidrule(lr){10-11}
    Layout & 1 & 2 & 1 & 2 & 1 & 2 & 1 & 2 & 1 & 2\\
    \midrule
    JUWELS  & 72\% & 65\% & 62\% & 60\% & 57\% & 57\% & 52\% & 54\% & 43\% & 44\%\\
    Ookami  & 27\% & 3\% & 22\% & 4\% & 20\% & 8\% & 19\% & 15\% & 12\% & 10\%\\
    HAICGU & 69\% & 62\% & 64\% & 54\% & 62\% & 53\% & 56\% & 54\% & 50\% & 50\%\\
    \midrule[\heavyrulewidth]
    \bottomrule
  \end{tabular}


\label{tab:archeff}
\end{table}

The single-node performance comparison is illustrated in Fig.~\ref{fig:comp} and the corresponding architectural efficiencies are listed in Table~\ref{tab:archeff}, revealing notable performance variations between testing platforms. 

The performance differences on JUWELS and HAICGU are roughly in line with the STREAM Triad bandwidth differences in Table~\ref{tab:spec}. Despite that the Ookami nodes featuring a 4$\times$ and 2.8$\times$ higher memory bandwidth than the JUWELS and HAICGU nodes, respectively, this benefit could not be exploited. One reason is that the number of cores on HAICGU is much larger than that on Ookami, enabling greater memory-level parallelism through higher thread utilization. Another reason is that CPUs on both JUWELS and HAICGU feature substantially larger cache capacities compared to the A64FX on Ookami. These architectural differences are reflected in the performance: HAICGU outperforms Ookami with rhs blocking (achieves up to 24\% of speedup compared to the unblocked implementation), while JUWELS reaches about 70\% of their performance.

\subsubsection{Performance Analysis}
\label{subsubsec:kernelanalysis}

Given the detailed performance results, we raise key questions regarding the effectiveness of our optimizations and the observed anomalies.

\noindent\textbf{How do different data layouts affect SIMD utilization, and how does this impact overall performance?}

We implement Layout 2 for efficient SIMD utilization as explained in Section~\ref{subsec:datalayout}, and keep the original Layout 1 as a legacy and a comparison baseline. To quantify SIMD utilization in rhs blocking, we use PAPI \cite{browne2000portable} to read performance counters relating to the executed SIMD instructions. 

For JUWELS, we measure 512-bit AVX instructions executed by reading the \texttt{FP\_ARITH:512B\_PACKED\_DOUBLE} event. Table~\ref{tab:fp512} reveals that Layout 2 executes more such instructions than Layout 1, except for $b=2$, where using 512-bit registers for vectorizing scalar-vector operations is less efficient due to the vector length of 2 and element sizes being 128-bit. Fig.~\ref{fig:dirac1} and~\ref{fig:comp} illustrate that Layout 2 outperforms Layout 1 for $b > 2$ on JUWELS, and this performance advantage is supported by its more effective SIMD utilization.


\begin{table}[btp]
    \centering
        \caption{\texttt{FP\_ARITH:512B\_PACKED\_DOUBLE} readings of the Wilson-Dirac operator evaluation for multiple rhs on JUWELS.}
    \begin{tabular}{l @{\hspace{1cm}}c @{\hspace{0.5cm}}c @{\hspace{0.5cm}}c@{\hspace{0.5cm}} c}  
\toprule
   $b$ = & 2  & 4 & 8 & 16\\
\midrule
Layout 1  & 22675456     & 45350912     &  90701824  & 181403648  \\
Layout 2   & 11075584     & 59768832     &  160890880 &  305463296\\
\midrule[\heavyrulewidth]
\bottomrule
\end{tabular}
\label{tab:fp512}
\end{table}


While both layouts utilize AVX512 on JUWELS, we observe that only Layout 2 utilizes SVE despite it being enabled for both layouts on Ookami. Therefore, we measure both SVE (\texttt{FP\_DP\_SCALE\_OPS\_SPEC}) and NEON (\texttt{FP\_DP\_FIXED\_OPS\_SPEC}) instructions for SIMD utilization analysis. Table~\ref{tab:armsimd} shows that SVE instructions are generated for Layout 2, whereas only 128-bit SIMD instructions are generated for Layout 1. However, Layout 2 generally underperforms compared to Layout 1 on Ookami (Fig.~\ref{fig:dirac1} and~\ref{fig:comp}), despite that fewer instructions are executed. This may indicate a higher cost of SVE instructions on Ookami.

\begin{table}[btp]
    \centering
        \caption{\texttt{FP\_DP\_SCALE\_OPS\_SPEC} (Scale) and \texttt{FP\_DP\_FIXED\_OPS\_SPEC} (Fixed) readings of the Wilson-Dirac operator evaluation for multiple rhs on Ookami.}
\begin{tabular}{l l*{4}{c}}
    \toprule
      $b=$ & & 2 & 4 
      & 8 & 16 \\
    \midrule
    \multirow{2}{*}{Scale} & Layout 1  & 0  & 0  & 0  & 0 \\
    & Layout 2  & 39322800  & 39322800  & 41420016 & 82840032 \\
    \multirow{2}{*}{Fixed} & Layout 1 & 45090149 & 90180149 & 180360293 & 360720437 \\
    & Layout 2 & 5  & 5  & 5  & 5 \\
    \midrule[\heavyrulewidth]
    \bottomrule
  \end{tabular}
\label{tab:armsimd}
\end{table}

Due to the lack of documentation on performance counters for the Kunpeng 920 processor, we cannot perform a quantitative SIMD analysis on HAICGU. However, we note that each core on HAICGU contains just one 128-bit double-precision SIMD pipeline, creating a performance bottleneck for SIMD operations. Consequently, Layout 1 exhibits a slightly better performance to Layout 2 in most cases on HAICGU (Fig.~\ref{fig:comp}).

\noindent\textbf{What can we tell about performance portability?}


As shown in Fig.~\ref{fig:dirac1} and Table~\ref{tab:archeff}, our implementation of the Wilson-Dirac operator evaluation achieves high architectural efficiency on both JUWELS and HAICGU, marking a successful porting of DD-$\alpha$AMG to Arm. Notably, Ookami demonstrates the lowest efficiency despite having the highest theoretical memory bandwidth, indicating that its architecture presents greater challenges for achieving full bandwidth saturation for the code. 


\noindent\textbf{Why does the performance not grow in proportion to its arithmetic intensity?}

For memory-bound applications, the roofline model suggests that the performance ceiling should grow in proportion to its arithmetic intensity. Assuming sustained memory bandwidth saturation, the measured performance is also expected to scale accordingly. However, the performance trends in Fig.~\ref{fig:dirac1} and~\ref{fig:comp} suggest that additional factors that are not captured by the roofline model influence the results, causing the curves to rise with a different gradient than expected and to bend downward for a large blocking size $b$. This behavior is consistent with the observed decreases in architectural efficiency.

We investigated several reasons for performance degradation. First, network bandwidth can become a bottleneck for a large $b$ as more data need to be communicated. Second, cache effects may come into play, as a large $b$ can introduce strides in data access. Third, the effective memory bandwidth may vary with $b$, e.g., due to changes in the read-to-write ratio. 

The first reason can be ruled out because Algorithm~\ref{alg:wilson-dirac} effectively overlaps computation and communication. The communications in lines 3 and 5 are paired with lines 6 and 8, respectively. Sufficient computation in lines 4 and 7 minimizes the waiting time. In practice, we observe that MPI\_Wait() time is about 10\% of the computation time, independent of $b$.  This is also supported by the nearly identical trends in Fig.~\ref{fig:dirac1} and~\ref{fig:comp} for multi-node and single-node runs. 

Given the above, we continue our analysis focusing solely on single-node performance. For quantitative analysis, we use available performance counters on Ookami to evaluate the $64\times 16^4$ lattice with only MPI enabled (See Table~\ref{tab:cache}), eliminating threading interference with counter readings.

\begin{table*}[btp]
    \centering
        \caption{\texttt{L2D\_CACHE\_REFILL}, \texttt{L2D\_CACHE\_WB}, \texttt{CPU\_CYCLES} readings and derived bandwidth in the Wilson-Dirac operator evaluation for multiple rhs on Ookami.}
    \begin{tabular}{{l}*{10}{c}}  
\toprule
        $b=$& \multicolumn{2}{c}{1} & \multicolumn{2}{c}{2} & \multicolumn{2}{c}{4} & \multicolumn{2}{c}{8} & \multicolumn{2}{c}{16}\\
    \cmidrule(lr){2-3}
    \cmidrule(lr){4-5}
    \cmidrule(lr){6-7}
    \cmidrule(lr){8-9}
    \cmidrule(lr){10-11}
 Layout &  1 &  2 &  1  &  2 &  1  &  2 & 1  &  2 &  1  &  2 \\
    \midrule
Refill  & 392270 &  400478  & 581620 & 677120  &  979685 & 1238074 & 1745437 & 2365484 & 3272209 & 3422734 \\
Writeback  & 165508   & 164374  & 238375 & 325464     &  422127 & 652722 &  856033 & 1303890 & 1617233 & 1516062\\
Cycles ($10^6$)  & 32 & 357     & 63     &  361 & 125 &  311 & 244 & 305 & 490 & 619\\
Bandwidth(GB$s^{-1}$) & 127 & 12  & 97  & 21 & 83  & 45  & 79  & 89  & 74  & 59 \\
\midrule[\heavyrulewidth]
\bottomrule
\end{tabular}
\label{tab:cache}
\end{table*}

Unlike most stencil algorithms, Algorithm~\ref{alg:wilson-dirac} does not rely on cache reuse. It iterates over lattice sites for each $\mu$, the data are not reused within a fixed $\mu$ as the sites share no neighbors in the same direction. In the best-performing variant (Layout 1) on Ookami, cache refill and writeback scale linearly with $b$, indicating little change in temporal locality exploitation.

After excluding two possibilities, we turn to the analysis of the read-to-write ratio. The read-to-write ratio for the Wilson-Dirac operator evaluation described with Algorithm~\ref{alg:wilson-dirac} is $204b + 330:204b$, without accounting for data locality or cache write-allocate behavior. Given that the A64FX has a peak write throughput of \qty{461}{\giga\byte\per\second} (half its peak read throughput), the ratio suggests that the code may underutilize the read throughput and become more write-intensive as $b$ increases. For example, the cache refill to writeback ratio of Layout 1 decreases from around 2.4 at $b=1$ to 2 at $b=16$, as shown in Table~\ref{tab:cache}. For reference, the STREAM Copy benchmark reports a ratio of 2.43 with \qty{540}{\giga\byte\per\second} bandwidth, 13\% less than the Triad bandwidth on Ookami.


We further measure the bidirectional effective bandwidth (included in Table~\ref{tab:cache}) between L2 cache and memory with PAPI on Ookami using
\begin{equation} \label{bw:eq}
\frac{(\texttt{L2D\_CACHE\_REFILL} + \texttt{L2D\_CACHE\_WB}) \times \qty{256}{\byte} \times f}{\texttt{CPU\_CYCLES}},
\end{equation}
where \qty{256}{\byte} is the cache-line size and $f$ the processor frequency. We observe a continuous effective memory bandwidth decline for Layout 1, dropping 40\% from $b=1$ to $16$. Although the roofline model predicts a 60\% performance increase, the combined effect of these factors results in almost no net gain for rhs blocking. 


Although JUWELS lacks performance counters to track hardware prefetcher behavior, making it impossible to obtain precise memory load and store information, it can be inferred from the STREAM benchmark results that memory bandwidth decreases if the ratio of memory stores versus loads increases. For example, the STREAM Copy and Scale benchmarks achieve about \qty{133}{\giga\byte\per\second}, which is 14\% less than the Add and Triad bandwidths on JUWELS. Namely, changes in the read-to-write ratio are likely to impact performance on JUWELS, explaining non-proportional performance changes.


Comprehensive memory load and store monitoring could not be implemented on HAICGU because the lack of performance counter documentation. However, blocking size increases affect HAICGU's performance less significantly than the other platforms. For reference, the STREAM Copy and Scale benchmarks achieve about \qty{214}{\giga\byte\per\second} on HAICGU, which are nearly identical to its Add and Triad bandwidths. This suggests that read-to-write ratio affects HAICGU's performance substantially less than the other platforms.


\noindent\textbf{Are there additional factors that impact performance?}
We observe from the assembly code generated by GCC with auto-vectorization that different choices of blocking size $b$ can produce largely different code. These differences may stem from decisions regarding loop unrolling, SIMD instructions, and even the use of complex arithmetic. Taking a simple routine of vector-scalar operation (cf. Line 2 in Algorithm~\ref{alg:wilson-dirac}) in Layout 2 for example, the compiler on Ookami unrolls the loop over $b$, i.e., it performs vector-scalar multiplication for all rhs in one iteration. This decision leads to significantly more data load and store instructions for \(b=8,16\) compared to for \(b=1,2\). This suggests that the actual arithmetic intensity at the machine level may not scale as expected based on code analysis. Therefore, performance predictions using the roofline model with Equation~\ref{eq:arithmetic} might be overly optimistic. This also helps explain the observed non-proportional performance changes relative to the arithmetic intensity given by Equation~\ref{eq:arithmetic}.

\subsection{Batched GMRES Solver}
\label{subsec:bgmreseval}

\subsubsection{Multi-Node Performance}
To fully assess the impact of our optimizations to the Wilson-Dirac kernel, we now analyze the performance within an LQCD simulation framework involving an iterative solver. For this purpose, we evaluate the batched GMRES solver described in Algorithm~\ref{alg:bgmres}, and fixed the iteration count to 100 (10 restarts with length 10).

\begin{figure}[btp]
  \centering
  \includegraphics[width=0.46\textwidth]{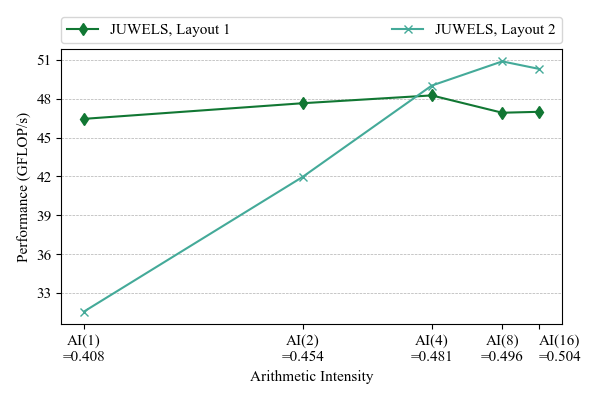}
  \vspace{-3mm}
  \caption{Performance per node scaling with arithmetic intensity for 100 iterations of batched GMRES for the $128\times64^3$ lattice on JUWELS.}
  \label{fig:bgmres1}
\end{figure}

 As illustrated in Fig.~\ref{fig:bgmres1}, the performance trends of the two layouts are smoother than those in Fig.~\ref{fig:dirac1}, because the arithmetic intensity decreases as additional data, such as test vectors $V$ and Hessenberg matrix $H$ are stored for each rhs in batched GMRES. This lower arithmetic intensity reduces the peak performance speedup to 7\% compared to the unblocked implementation on JUWELS.



The performance of the odd-even preconditioning case is further constrained by increased non-sequential memory access patterns resulting from the evaluation of the Schur complement. We observe the performance of batched GMRES with odd-even preconditioning is approximately 70\% of that without preconditioning. The results reveal a trade-off between algorithmic efficiency and machine performance. While the machine performance is lower, the total iteration count to convergence is reduced by half for the odd-even preconditioned linear system, making it nevertheless more time-efficient. 

The batched GMRES structure, inherited from the legacy DD-$\alpha$AMG code, is single-threaded and suffers from inefficient memory bandwidth utilization on Ookami. It does not show an advantage over that on JUWELS. Hence, we omit the discussions of batched GMRES on Ookami here. 

\subsubsection{Performance Analysis}

Since the Wilson-Dirac operator evaluation kernel takes over 90\% of computation time in batched GMRES, the performance trend of batched GMRES is very similar to that of the kernel. Namely, the causes of performance deviations with respect to data layouts and the roofline model are explained in Section~\ref{subsubsec:kernelanalysis}. An additional abnormality occurs in the performance of batched GMRES for $b\leq4$ of Layout 2 in Fig.~\ref{fig:bgmres1}, where the performance improvement surpasses the growth seen in the kernel in Fig.~\ref{fig:dirac1}. At least in parts, this behavior is caused by the dot product operations in Arnoldi's method, which become notably less efficient for Layout 2 when the blocking size is small. 

The dot product is computed between an Arnoldi vector $w$ and the output vector $\eta$ from the Wilson-Dirac operator evaluation; both have the form given in Equation~\ref{eq:block}. The operation to be performed is $h^{(i)}=w^{(i)}\cdot\eta^{(i)}$ for $i=1\ldots b$. 

\begin{figure}[btp]
\vspace{-2.5mm}
    \centering
    \subfloat[Layout 1]{%
        \includegraphics[width=0.235\textwidth]{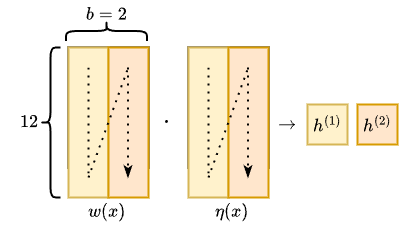}
    }
    \hfill
    \subfloat[Layout 2]{%
        \includegraphics[width=0.235\textwidth]{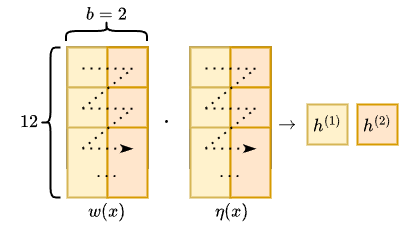}
    }
    \caption{Example of dot product operation using two data layouts with $b=2$, where colors distinguish separate RHS vectors.}
    \label{fig:dotprod}
\end{figure}

To showcase the dot product computation, we illustrate $h^{(i)} = w^{(i)}(x) \cdot \eta^{(i)}(x)$ in Fig.~\ref{fig:dotprod}. Consider that 512-bit registers are used, a load of $w(x)$ and $\eta(x)$ for Layout 1 results in sequential data corresponding to either the first or second rhs, a dot product of them contributes to either scalar $h^{(1)}$ or $h^{(2)}$, respectively. However, the data are interleaved in Layout 2, a dot product using 4 elements of $w(x)$ and 4 elements of $\eta(x)$ contributes to both $h^{(1)}$ or $h^{(2)}$, requiring additional operations to separate the data from the 512-bit registers for reduction. 

We introduced a fix to delay data separation until the final step, and used a buffer to store additional data for small $b$. This approach improves performance by 4–5\% for $b=1,2$, which tends to align the performance growth of Layout 2 in batched GMRES with the kernel. 



%% file: 06_SME.tex
\section{Emulated Performance Evaluation of SME}
\label{section:smeanalysis}



Using the instruction-cost evaluation method specified in Section~\ref{subsubsec:emu}, we count the FP, NEON, SVE, and SME instructions (including loads and stores) in the experiments, assuming that integer instructions have negligible execution costs. The experiments are conducted for the blocking sizes 8 and 16, with identical results in both cases. Detailed numbers are provided in Table~\ref{tab:sme}. 

Without cost overrides, where each instruction is weighted as 1, the best SME variant neg-A achieves a 2.73$\times$ speedup over SVE and 12.2$\times$ over BLIS. The neg-M variant, though slowed by additional \verb|revd| and \verb|fneg| instructions for large $b$s as discussed in Section~\ref{subsec:smekernel},  still maintains a 42\% advantage over SVE and a 6.3$\times$ improvement versus BLIS.

\begin{table}[btp]
    \centering
        \caption{Instruction cost for \qty{512}{\bit} SVL with $b=8,16$ and iteration count 100.}
         \begin{tabular}{l@{\hspace{1cm}}c @{\hspace{1cm}} c @{\hspace{1cm}} c @{\hspace{1cm}} c}
    \toprule
     Method & BLIS & SVE & Neg-M & Neg-A  \\
    \midrule
    Override OFF  & 31704 & 7104 & 5004 & 2604\\
    Override ON  & 29122 & 8022 & 6222 & 2922 \\
    \midrule[\heavyrulewidth]
    \bottomrule
    \label{tab:sme}
    
    
    \end{tabular}
\label{tab:cost}
\vspace*{-5mm}
\end{table}

We also investigate the cost with the overrides, such that \verb|mov| instructions cost 0 (assuming it is a register rename), store costs 2, and SME outer-product instructions cost 2. Doubling store costs over loads aligns with A64FX's architecture, which has twice as many load ports as store ports. Similarly, doubling SME outer-product costs reflects their lower throughput compared to simpler arithmetic operations. With these overrides, the SME variant neg-A achieves a 2.75$\times$ speedup over SVE and 9.97$\times$ over BLIS. The neg-M variant outperforms SVE by 30\% and BLIS by 4.7$\times$.

%% file: 07_conclusions.tex
\section{Conclusions}
\label{section:conclusions}

We present several optimizations for the LQCD solver DD-$\alpha$AMG. Firstly, we discuss implementations of rhs blocking in the Wilson-Dirac operator evaluation and the batched GMRES solver. Secondly, we introduce a data layout designed to enhance SIMD utilization for rhs blocking in the Wilson-Dirac kernel. Finally, we implement an LQCD routine with SME. The artifacts for this paper, including all source code, scripts, and datasets necessary to reproduce our results, are available online~\cite{shiting_long_2025_17486533}.

We empirically evaluate our optimizations of the kernel and the solver on JUWELS, Ookami, and HAICGU, and analyze performance portability across the platforms. Our performance analysis indicates that increasing the blocking size generally improves performance; however, this also leads to higher write intensity, which can offset gains and potentially cause performance degradation. Additionally, compiler behavior can further impact performance in unpredictable ways. We conclude that while using multiple rhs is a common strategy to accelerate memory-bound scientific applications, achieving optimal performance requires careful consideration of both hardware and compiler constraints. It also involves trade-offs when algorithmic changes are applied, such as shown in the evaluation of the batched GMRES solver, which may lead to reduced machine performance. 

We use an instruction cost model to evaluate emulated performance of the LQCD routine implemented using SME instructions, which demonstrates a notable advantage over that of other implementations. Exploiting this performance requires, however, a sufficiently high memory bandwidth.

Future work includes optimizing data locality in the Wilson-Dirac operator evaluation. Intermediate values could be reused immediately with modifications to the loop structures. Such a change also addresses the write-intensive behavior observed with large blocking sizes. It would also be valuable to implement critical parts of the kernel using intrinsics and compare the results with those presented in this paper. Moreover, the interplay between scalability and performance is an important aspect to investigate. If HPC-grade SME hardware becomes available, we plan to test our implementations in real-world scenarios for validation.
